\documentclass{article}
\usepackage{spconf,amsmath,graphicx}
\usepackage[linesnumbered,ruled,vlined]{algorithm2e}
\usepackage{float}
\usepackage{multirow}
\usepackage[table]{xcolor}

\title{MedMAE: A Self-Supervised Backbone for Medical Imaging Tasks}
\name{Anubhav Gupta, Islam Osman, Mohamed S. Shehata, and John W. Braun}
\address{University of British Columbia\\
3333 University Way, Kelowna, BC}
\begin{document}
%
\maketitle
\begin{abstract}
Medical imaging tasks are very challenging due to the lack of publicly available labeled datasets. Hence, it is difficult to achieve high performance with existing deep-learning models as they require a massive labeled dataset to be trained effectively. An alternative solution is to use pre-trained models and fine-tune them using the medical imaging dataset. However, all existing models are pre-trained using natural images, which is a completely different domain from that of medical imaging, which leads to poor performance due to domain shift. To overcome these problems, we propose a large-scale unlabeled dataset of medical images and a backbone pre-trained using the proposed dataset with a self-supervised learning technique called Masked autoencoder. This backbone can be used as a pre-trained model for any medical imaging task, as it is trained to learn a visual representation of different types of medical images. To evaluate the performance of the proposed backbone, we used four different medical imaging tasks. The results are compared with existing pre-trained models. These experiments show the superiority of our proposed backbone in medical imaging tasks.
\end{abstract}
\section{Introduction}
\label{sec:intro}

Medical imaging tasks are very challenging due to the lack of publicly available labeled datasets. Hence, it is difficult to achieve high performance with existing deep-learning models as they require a massive labeled dataset to be trained effectively. An alternative solution is to use pre-trained models and fine-tune them using the medical imaging dataset. However, all existing models are pre-trained using natural images, which is a completely different domain from that of medical imaging, which leads to poor performance due to domain shift.

In this paper, we propose a large-scale unlabeled dataset of medical images
from different sources and ViT-based \cite{dosovitskiy2020image} backbone for any medical imaging task. The unlabeled dataset consists of images captured from different sources such as MRI, CT, and X-ray. These images are captured for different body parts such as the chest, lungs, pancreas, abdomen, lung, brain, and pelvis. The main goal of the collected dataset is to be extensive and diverse to allow the proposed model to gain useful knowledge of different types of medical images. Hence, this knowledge can be used to achieve high accuracy in any specific medical imaging task. The proposed model is ViT trained with the masked autoencoder (MAE) \cite{he2022masked} technique. MAE is a self-supervised learning (SSL) technique that leverages the usage of unlabeled data, allowing the model to discover intrinsic features in the medical images. The ultimate objective of this paper is to propose a deep-learning model that learns versatile representations of medical images that can be effectively transferred to any medical imaging task, even when labels are scarce. The contribution of this paper is summarized as follows:
\begin{itemize}
\item Proposing a large-scale unlabeled medical imaging dataset that can be used for self-supervised and unsupervised learning techniques.
\item Proposing a Medical Masked autoencoder (MedMAE), a pre-trained backbone that can be used for any medical imaging task.
\end{itemize}

The rest of the paper is organized as follows: Section 2 provides a literature review of medical imaging models. Section 3 Shows the details of the proposed work. Section 4 depicts the results of the experiments that were conducted. Finally, Section 5 concludes and summarizes the paper and discusses the future directions.

\section{Related work}
SSL has emerged as a popular learning methodology in medical image analysis, particularly beneficial in scenarios where annotated data is scarce and there is an abundance of unlabeled data \cite{huang2023self}. Several researchers have
demonstrated the effectiveness of the SSL approach throughout various medical image analysis tasks such as detection and classification \cite{sriram2021covid},\cite{lu2020semi},\cite{li2021rotation}, detection and localization \cite{chen2019self},\cite{nguyen2020self},\cite{sowrirajan2021moco}, and segmentation tasks \cite{karani2020contrastive},\cite{taleb20203d},\cite{xie2020pgl}. Huang et al. provide an extensive overview of DL approaches employing SSL for medical image classification in \cite{huang2023self}. Our work primarily concentrates on self-prediction methods, which aligns with our strategy to create an SSL-based backbone tailored for medical imaging applications. Self-prediction involves augmenting or masking certain segments of an image, followed by an attempt to reconstruct the original image using the remaining unmasked parts. Several studies demonstrate the use of self-prediction methods for medical image restoration tasks. Jung et al. introduced a masked auto-encoder technique for functional connectivity matrix restoration in rs-fMRI data. The method involves masking different rows and columns of a matrix and then reconstructing the original matrix, where the functional connectivity matrix represents the relationship between different regions of interest for each subject \cite{jung2021inter}. Jana et al. proposed an encoder-decoder architecture tailored for restoring CT scans that have been corrupted by swapping several small patches within a single CT slice \cite{jana2021liver}. Liu et al. developed a U-net model for restoring ultrasound images that have been augmented through local-pixel shuffling. Clinical variables such as age, gender, and tumor size are then concatenated with the encoder’s outputs for downstream prediction tasks \cite{liu2021tn}. At present, there seems to be an absence of an SSL-based backbone dedicated to medical image understanding tasks. Recognizing the problems posed by the need for extensively annotated datasets, SSL emerges as a promising alternative capable of mitigating these challenges. Unlike supervised learning methods that typically require large quantities of labeled data, SSL has the potential to generate versatile models. These generalist models can be subsequently fine-tuned for a variety of downstream tasks, even in the absence of large labeled datasets.

\begin{table*}
    \centering
\resizebox{0.9\textwidth}{!}{
    \begin{tabular}{l|c|c|c}

    \hline
        Collection & Location & Subject & DataTypes\\ \hline
        NIH Chest X-Ray \cite{wang2017chestx} & Frontal Chest & 30805 & X-Ray \\
        Pseudo-PHI-DICOM-Data \cite{rutherford2021dicom} & Various & 21 & CR, CT, DX, MG, MR, PT\\
        COVID-19-NY-SBU \cite{saltz2021stony} & Lung & 1384 & CR, CT, DX, MR, PT, NM, OT, SR\\
        CTPred-Sunitinib-panNet \cite{chen2023special} & Pancreas & 38 & CT\\
        Stagell-Colorectal-CT \cite{li2023special} & Abdomen, Pelvis & 230 & CT\\
        CT images in COVID19 \cite{an2020ct} & Lung & 661 & CT\\
        MIDR-RICORD-1a\&1b \cite{tsai2021data} & Lung & 227 & CT\\
        CT-ORG \cite{rister2020ct} & Bladder, Brain, Kidney, Liver & 140 & CT\\
        Pelvic-Reference-Data \cite{yorke2019pelvic} & Pelvis, Prostate, Anus & 58 & CT\\
        QIN Lung CT \cite{kalpathy2015qin} & Lung & 47 & CT\\
        Pancreas CT \cite{roth2016data} & Pancreas & 82 & CT\\
        LungCT-Diagnosis \cite{grove2015quantitative} & Lung & 61 & CT\\
        NSCLC-Radiomics-Genomics \cite{aerts2015data} & Lung & 89 & CT\\
        RIDER Lung CT \cite{teng2023improving} & Chest & 32 & CT, CR, DX\\
        CT Colon ACRIN 6664 \cite{smith2015data} & Colon & 825 & CT\\
        LIDC-IDRI \cite{armato2011lung} & Chest & 1010 & CT, CR, DX\\
        TCGA-BLCA \cite{machado2024association} & Bladder & 120 & CT, CR, MR, PT, DX, Pathology\\
        TCGA-UCEC \cite{albertina2016cancer} & Uterus & 65 & CT, CR, MR, PT, Pathology\\
        COVID-19-AR \cite{desai2020data} & Lung & 105 & CT, DX, CR\\
        CMB-LCA \cite{biobank2022cancer} & Lung & 10 & CT, DX, MR, NM, US\\
        CMB-CRC \cite{biobank2022crccancer} & Colon & 12 & CT, DX, MR, PT, US\\
        TCGA-KIRC \cite{akin2016cancer}& Kidney & 267 & CT, MR, CR, Pathology\\
        CMB-PCA \cite{biobank2022pcacancer} & Prostate & 3 & CT, MR, NM\\
        CPTAC-CCRCC \cite{clark2013cancer}& Kidney & 222 & CT, MR, Pathology\\
        TCGA-SARC \cite{roche2016cancer} & Chest, Abdomen, Pelvis, Leg & 5 & CT, MR, Pathology\\
        TCGA-READ \cite{albertina2016radiology}& Rectum & 3 & CT, MR, Pathology\\
        TCGA-OV \cite{holback2016cancer} & Ovary & 143 & CT. MR, PAthology\\ \hline
    \end{tabular}
}
    \caption{A detailed overview of the various datasets collected to form MID the medical imaging dataset.}
    \label{tab:data}
\end{table*}

\section{Methodology}
We propose a large-scale unlabeled dataset of medical images and a backbone pre-trained using the proposed dataset with a self-supervised learning technique called Masked autoencoder. This backbone can be used as a pre-trained model for any medical imaging task, as it is trained to learn a visual representation of different types of medical images.

\subsection{Medical Imaging Dataset}
Our primary goal is to pre-train MedMAE-B on a large and diverse dataset of medical images. This pre-training aims to achieve two key objectives: 1) Capture Latent Representations: By processing a vast array of medical images, MedMAE-B will learn underlying characteristics specific to this domain. This translates to a model with a deeper grasp of medical image content. 2) Boost Generalizability and Versatility: The pre-training process exposes the model to diverse medical scenarios. This, in turn, enhances its ability to adapt and perform well on various downstream tasks related to medical image analysis.

To achieve effective pre-training, we constructed a massive dataset encompassing various body parts and image types. This dataset integrates images from multiple repositories and platforms (details in Table \ref{tab:data}). We meticulously documented all data sources to ensure transparency. We call the called dataset MID (Medical Imaging Dataset).
Each image within the compiled dataset underwent specific transformations before feeding into the MAE architecture. This included: 1) Converting all files to a compatible image format. 2) Resizing images to meet the model's input specifications. 3) Removing corrupted or erroneous files.
We opted against data augmentation techniques to artificially inflate dataset size. Instead, we focused on the inherent diversity within the vast collection of datasets.

The current dataset comprises over 2 million images across various medical imaging modalities (CT scans, X-rays, MRI, etc.). This extensive and ever-growing dataset equips MedMAE-B with a robust and versatile understanding of medical imagery.

\begin{figure*}
\centering
\includegraphics[height=8.cm]{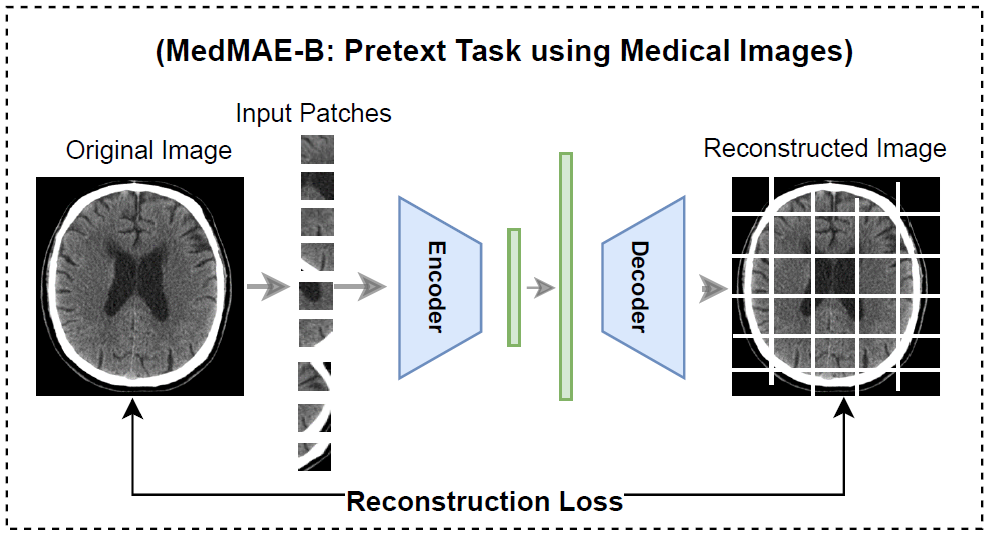}
\caption{MedMAE architecture: The process is initiated by randomly
masking 75\% of the original image and inputting the remaining 25\% of visible patches into the encoder, which captures the latent representations and encodes the patches. Subsequently, the aim of the decoder is to reconstruct the complete image using the encoded and masked patches. The reconstruction loss helps to improve the reconstruction with each iteration.}
\label{fig:feshnet}
\end{figure*}

\subsection{MedMAE Architecture in pre-training}
This section delves into the core building blocks of Medical Masked Autoencoders (MedMAE): the encoder, decoder, and loss function. 

\textbf{Encoder} The MedMAE encoder leverages a Vision Transformer (ViT) architecture. The input image/volume is first divided into non-overlapping patches. These patches are then randomly assigned into two groups: visible and masked patches. The encoder operates solely on the visible patches, aiming to learn a meaningful representation based on this partial information.
To compensate for the lack of complete spatial information, each patch is augmented with a corresponding positional embedding before being fed into the ViT. This positional encoding helps the encoder maintain an understanding of the relative location of each patch within the larger image/volume.  Crucially, since the encoder's output is used to reconstruct the masked regions, it is incentivized to extract a comprehensive representation from these incomplete observations. 

\textbf{Decoder} The goal of the decoder is to fill the gaps (i.e., predict the masked patches). The MedMAE decoder receives two sets of tokens as input: 1) Patch-wise representations: These are the outputs generated by the encoder for the visible patches. 2) Mask tokens: These are learnable tokens representing the masked regions. They are inserted into the decoder at the corresponding positions where patches are masked in the original input. Similar to the encoder, positional embeddings are added to all input tokens in the decoder. This enables the decoder to reconstruct the missing information at each masked position.  It's important to note that the decoder serves as an auxiliary module used only during the pre-training stage. It is not employed in downstream tasks where the pre-trained encoder plays a central role.

\textbf{Loss function} MedMAE leverages a reconstruction loss function, specifically the mean squared error (MSE). However, unlike traditional autoencoders that aim to reconstruct the entire input, MedMAE only focuses on predicting the pixel values of the masked patches. This approach has been shown to yield superior results.  In practice, for better training stability, the normalized pixel values within each masked patch are used as reconstruction targets instead of the raw values.

\subsection{MedMAE Architecture in downstream tasks}
After MedMAE is pre-trained, for each downstream task, we replace the MedMAE decoder with a task-specific head. In the classification task, the task-specific head is simply a fully connected layer with the number of neurons equal to the number of classes. This layer is followed by a softmax activation function. On the other hand, for the segmentation task, four transposed convolutional layers are added to scale up the dimensions of the embeddings of the MedMAE encoder, followed by a convolutional layer with a single filter to produce the output mask. This layer is followed by a sigmoid activation function. During the training of the downstream tasks, only the parameters of the added head are updated. A well-known process which is called linear probing.

\section{Experiments and Results}

\subsection{Implementation details}
In this section, we delve into the outcomes of our MedMAE-B model, which underwent an extensive pre-training process spanning over 1000 epochs. Each epoch took approximately $1$ hour and $20$ minutes to train. Therefore, the model has a training time of more than $1000$ hours in total. As previously stated, our evaluation was exclusively based on the MAE ViT-B model. We configured our model with a batch size of $64$, and the dimensions of the input images were set to $224\times224$ pixels. The transformation and normalization of the entire image dataset were facilitated through the transform function. It is crucial to highlight that our study solely focuses on grayscale medical images without incorporating any colored image data. To set the learning parameters, we established a base learning rate of $10^{-3}$.

\subsection{MedMAE evaluation}
To evaluate MedMAE, we conducted four experiments. Two experiments with private datasets and two experiments with publicly available datasets. The first experiment is Automating quality control for both CT scanners and MRI scanners. To ensure the reliability of radiology scanners, quality assurance protocols are routinely implemented, typically on a daily or weekly basis \cite{mcrobbie2017quality}. It is worth noting that these procedures are time-consuming and necessitate taking the scanner offline during the assessment process. The goal is to automate this process using deep-learning to detect whether an image is captured by a well-calibrated or a miss-calibrated scanner. In this experiment, we have a dataset collected from two hospitals for CT and MRI scans. The images in the dataset are labeled with either pass or fail. Where pass means the image is captured by a well-calibrated CT scanner. While fail means the image is captured by a miss-calibrated CT scanner. In the second experiment, we have images captured for the breast, and the goal is to detect if there is a disease or not. For the third experiment, we used the publicly available dataset CVC-ClinicDB for a segmentation task. Finally, the last experiment is pneumonia detection using the publicly available dataset ChestX-ray14. We compared the results of our proposed model with existing models pre-trained on the ImageNet (IN1k) dataset \cite{deng2009imagenet}. We chose five different models that have almost the same number of parameters. These models are: ResNet \cite{he2016deep}, EfficientNetv2-S \cite{tan2021efficientnetv2}, ConvNext-B \cite{liu2022convnet}, ViT-B \cite{dosovitskiy2020image}, and Swin-B \cite{liu2021swin}. All these models are pre-trained with supervised learning on ImageNet. Additionally, we compared our results with the original MAE \cite{he2022masked}. Evaluation is done by linear probing all models for 300 epochs on the training set and performance on the testing set is reported for each experiment. Linear probing is the process of fine-tuning only the last fully connected layer of the model while keeping the rest of the model parameters unchanged (i.e., pre-trained parameters).

\subsubsection{Task1: Automating quality control for CT and MRI scanners}
The precision and promptness of patient diagnoses conducted on radiology scanners are inherently linked to the quality of the images generated during the scanning process. When the quality of these images is compromised, it can have dire consequences. An image obtained from a miss-calibrated scanner can necessitate a costly and time-consuming repetition of the patient scan. In more critical scenarios, if such issues go undetected, they can give rise to missed diagnoses, posing substantial risks to patient health and, in certain instances, even resulting in tragic outcomes, including loss of life \cite{dehkordi2019influence}, \cite{parrish2000impact}, \cite{song20193d}, and \cite{rubenstein1997image}. Providing real-time QC of the scanner calibration to radiologists can have several benefits: 1) improved patient throughput and reduced wait times, primarily by minimizing scanner downtime, and 2) enhanced diagnostic accuracy by swiftly identifying and addressing any image quality issues.

\textbf{CT Scan Dataset}
Specialized datasets were built to meet the unique requirements of this experiment. We procured patient studies. These studies were retrieved in the form of DICOM files. A meticulous de-identification process was applied to each scan, ensuring the absence of any identifiable patient data. After that, each DICOM file was converted to an image format. The images represent different body parts captured by a CT scanner. Each image is labeled with either pass or fail. Hence, the task to be performed is binary image classification. The collected data comprises CT scan images of different body parts for more than 100 unique patients, with a total number of images of more than 30000.

\begin{table}[]
\centering
\begin{tabular}{c|lll}
\hline
Pre-training                                                                 & Dataset & Backbone       & Accuracy \\ \hline
\multirow{5}{*}{Supervised}                                                 & IN1k    & ResNet         & 75.6     \\
                                                                            & IN1k    & EfficientNet-S & 71.3     \\
                                                                            & IN1k    & ConvNext-B     & 76.8     \\
                                                                            & IN1k    & ViT-B          & 78.2     \\
                                                                            & IN1k    & Swin-B         & 77.5     \\ \hline
Self- & IN1k    & MAE            & 78.5     \\
\rowcolor{gray!20}
     supervised                                                                       & MID     & MedMAE (Ours)  & \textbf{90.2}     \\ \hline
\end{tabular}
\caption{Results of our proposed model MedMAE against other models on CT dataset. The reported accuracy is the accuracy of linear probing pre-trained models.}
\label{tab:ct}
\end{table}

Table \ref{tab:ct} shows the results of our proposed model in comparison with existing deep-learning models. As shown in Table \ref{tab:ct}, the performance of our proposed model is superior to other models. Even in comparison with the same model but pre-trained on Imagenet (i.e., MAE), there is still a huge gap in performance of approximately $12\%$. It is important to mention that we could not pre-train other models using our dataset because these models are trained using supervised learning and our dataset has no labels.

\textbf{MRI Scan Dataset}
This dataset is collected the same way as the CT scan dataset but the images are captured from an MRI scanner instead of a CT. The dataset consists of images of three body parts: abdominal, head, and shoulder. The collected number of cases (patients) is 426. Each case has around 20 images. Each image is labeled with either pass or fail, the same as the CT dataset.

\begin{table}[]
\centering
\begin{tabular}{c|lll}
\hline
Pre-training                                                                 & Dataset & Backbone       & Accuracy \\ \hline
\multirow{5}{*}{Supervised}                                                 & IN1k    & ResNet         & 71.6     \\
                                                                            & IN1k    & EfficientNet-S & 66.1     \\
                                                                            & IN1k    & ConvNext-B     & 71.9     \\
                                                                            & IN1k    & ViT-B          & 72.7     \\
                                                                            & IN1k    & Swin-B         & 72.1     \\ \hline
Self- & IN1k    & MAE            & 74.3     \\

\rowcolor{gray!20}            supervised                                                                & MID     & MedMAE (Ours)  & \textbf{85.6}     \\ \hline
\end{tabular}
\caption{Results of our proposed model MedMAE against other models on MRI dataset. The reported accuracy is the accuracy of linear probing pre-trained models.}
\label{tab:mri}
\end{table}

Table \ref{tab:mri} shows the results of our proposed model against other models. Our model MedMAE shows a superior performance with a gap of $11.3\%$. It is important to mention that both the MRI and CT scan datasets are not shown during the pre-training phase. However, the pre-training dataset MID has images of MRI and CT collected from other datasets.

\subsubsection{Breast Cancer Prediction}
This dataset contains data from 29,686 patients. Each patient has multiple CT studies collected from 1988-2018. The dataset contains more than 5 million CT images of the chest area. We split the dataset into training, validation, and testing sets with ratios of $70\%$, $15\%$, and $15\%$, respectively. The results are shown in Table \ref{tab:bc}. MedMAE has a performance gain of $8.9\%$ over the MAE pre-trained on ImageNet. 

\begin{table}[]
\centering
\begin{tabular}{c|lll}
\hline
Pre-training                                                                 & Dataset & Backbone       & Accuracy \\ \hline
\multirow{5}{*}{Supervised}                                                 & IN1k    & ResNet         & 79.9     \\
                                                                            & IN1k    & EfficientNet-S & 75.1     \\
                                                                            & IN1k    & ConvNext-B     & 83.8     \\
                                                                            & IN1k    & ViT-B          & 84.0     \\
                                                                            & IN1k    & Swin-B         & 81.3     \\ \hline
Self- & IN1k    & MAE            & 84.3     \\

\rowcolor{gray!20}   supervised                                                                         & MID     & MedMAE (Ours)  & \textbf{93.2}     \\ \hline
\end{tabular}
\caption{Results of our proposed model MedMAE against other models on breast cancer dataset. The reported accuracy is the accuracy of linear probing pre-trained models.}
\label{tab:bc}
\end{table}

\subsubsection{Pneumonia detection}
We used the ChestX-ray14 \cite{wang2017chestx} dataset in this experiment. This dataset comprises 112120 frontal-view X-ray images of more than 30000 unique patients. For this dataset, we report the results of our model MedMAE against the same models from previous experiments. Additionally, we report the results of MedMAE against state-of-the-art models in this specific dataset. Table \ref{tab:chestx} shows the results of the linear probing of our models and other selected models on the ChestX-ray14 dataset. As shown in the table, MedMAE has $2.2\%$ accuracy more than the best model. 
In Table \ref{tab:chestxsota}, to have a fair comparison with other models, we fine-tuned all parameters of MedMAE instead of doing linear probing. This is due to the fact the state-of-the-art models are trained using supervised learning using this dataset (i.e., all model parameters are updated). As shown in the table, our model MedMAE shows a significant improvement in performance in comparison with other models, with a performance gain of $3.6\%$. 
\begin{table}[]
\centering
\begin{tabular}{c|lll}
\hline
Pre-training                                                                 & Dataset & Backbone       & Accuracy \\ \hline
\multirow{5}{*}{Supervised}                                                 & IN1k    & ResNet         & 66.4     \\
                                                                            & IN1k    & EfficientNet-S & 62.9     \\
                                                                            & IN1k    & ConvNext-B     & 67.5     \\
                                                                            & IN1k    & ViT-B          & 67.8     \\
                                                                            & IN1k    & Swin-B         & 67.1     \\ \hline
Self- & IN1k    & MAE            & 67.9     \\
\rowcolor{gray!20} supervised
                                                                            & MID     & MedMAE (Ours)  & \textbf{70.1}     \\ \hline
\end{tabular}
\caption{Results of our proposed model MedMAE against other models on ChestX-ray14 dataset. The reported accuracy is the accuracy of linear probing pre-trained models.}
\label{tab:chestx}
\end{table}

\begin{table}[]
\centering
\begin{tabular}{l|l}
\hline
Method             & Accuracy \\ \hline
Wang et al. \cite{wang2017chestx} & 73.8     \\
Yao et al. \cite{yao2017learning}        & 79.8     \\
CheXNet  \cite{rajpurkar2017chexnet}          & 84.4     \\
Google AutoML \cite{blog2017automl}     & 79.7     \\
NSGANetV1  \cite{lu2020multiobjective}        & 84.7     \\
MUXNet-m  \cite{lu2020muxconv}         & 84.1     \\
AE-CNN \cite{ranjan2018jointly}            & 82.41    \\
LEAF   \cite{liang2019evolutionary}            & 84.3     \\ \hline

\rowcolor{gray!20}
MedMAE (Ours)      & \textbf{88.0}  \\\hline  
\end{tabular}
\caption{Results of our proposed model MedMAE against state-of-the-art models on ChestX-ray14 dataset.}
\label{tab:chestxsota}
\end{table}

\subsubsection{Medical Segmentation}
In this experiment, we used the dataset CVC-ClinicDB \cite{bernal2015wm}. It is a dataset of frames extracted from colonoscopy videos. This dataset contains several examples of polyp frames and corresponding ground truth for them. The number of labeled images in this dataset is 612 images. The dataset is split into training, validation, and testing sets with ratios of $70\%$, $15\%$, and $15\%$, respectively. The split is done randomly. The task performed in this dataset is different from other datasets. In this dataset, our goal is to segment the region of interest and produce a binary mask, where the white pixels represent the region of interest and the black pixels represents everything else. In other experiments, we added a fully connected layer for the linear probing. However, in this experiment, we added a sequence of transposed convolutional layers to upscale the embeddings produced by the encoder to a resolution equivalent to the input image. The sequence of transposed convolutional layers consists of four layers with each layer followed by a GeLU \cite{hendrycks2016gaussian} activation function. The number of filters used in these layers is 256, 128, 64, and 32, respectively. All layers have a kernel size of $2\times2$ and stride of 2 (i.e., each layer increases the dimensions of the input embedding to double its size). After the sequence of layers, a final convolutional layer is added with a single filter of size $1\times1$ followed by a Sigmoid activation function to produce the output mask. As shown in Table \ref{tab:cvc}, our proposed model MedMAE outperforms other models by a noticeable gap of $6.8\%$ in the f-score. The f-score is calculated as follows:
\begin{equation}
    \mathcal{F} = \frac{2tp}{2tp+fp+fn}
\end{equation}
where $\mathcal{F}$ is the f-score, $tp$ is the true positive, $fp$ is the false positive, and $fn$ is the false negative.

\begin{table}[]
\centering
\begin{tabular}{c|lll}
\hline
Pre-training                                                                 & Dataset & Backbone       & $\mathcal{F}$ \\ \hline
\multirow{5}{*}{Supervised}                                                 & IN1k    & ResNet         & 57.9     \\
                                                                            & IN1k    & EfficientNet-S & 53.1     \\
                                                                            & IN1k    & ConvNext-B     & 61.8     \\
                                                                            & IN1k    & ViT-B          & 63.5     \\
                                                                            & IN1k    & Swin-B         & 60.2     \\ \hline
Self- & IN1k    & MAE            & 64.6     \\
\rowcolor{gray!20} supervised
                                                                            & MID     & MedMAE (Ours)  & \textbf{71.4}     \\ \hline
\end{tabular}
\caption{Results of our proposed model MedMAE against other models on CVC-ClinicDB dataset. The reported numbers are the f-score of linear probing pre-trained models.}
\label{tab:cvc}
\end{table}

\subsection{Visual results}
\begin{figure}[H]
\centering
\includegraphics[width=0.5\textwidth]{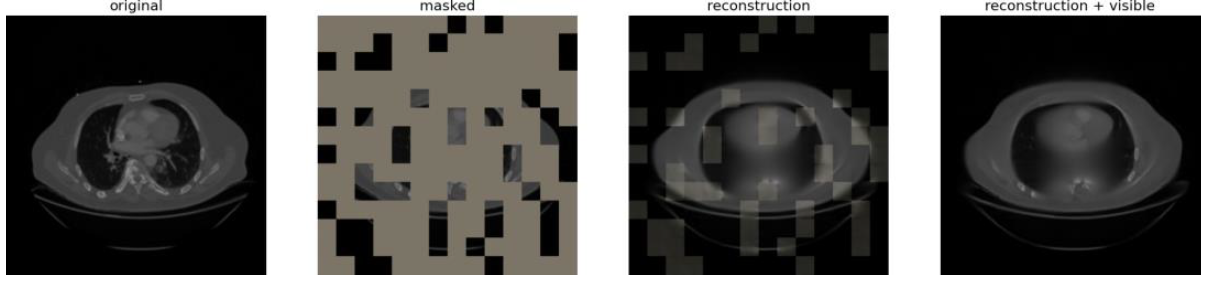}
\caption{Image construction using pre-trained MAE with natural images.}
\label{fig:vis1}
\end{figure}

\begin{figure}[H]
\centering
\includegraphics[width=0.5\textwidth]{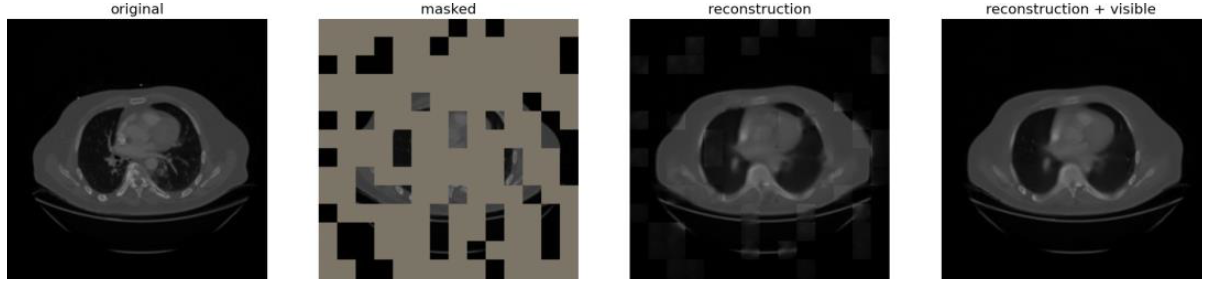}
\caption{Image construction using our proposed MedMAE.}
\label{fig:vis2}
\end{figure}

To offer a deeper understanding of the impact of the training process
on the model’s performance, we present examples of image reconstructions
both before and after the extensive training period. Figures \ref{fig:vis1} and \ref{fig:vis2} are examples of reconstruction results of a sample image taken from a dataset that was not utilized during the training process. Figures \ref{fig:vis1} and \ref{fig:vis2} present the comparison between the original and reconstructed image. This comparison highlights the initial performance of our MedMAE-B model in understanding and recreating the medical image. The refinement in the reconstructed image, when compared to earlier checkpoints, is clearly visible. These examples underscore the effectiveness of our extensive training process in enhancing the model’s proficiency in medical image analysis.

\section{Conclusion and Future Work}
In this paper, we proposed a large-scale unlabeled medical imaging dataset that has extensive and diverse medical images. Additionally, we proposed a ViT-based backbone that is pre-trained using the proposed dataset and we showed that this backbone can be used for various medical imaging tasks. We evaluated the proposed model using four tasks. The results of all tasks showed the superiority of our proposed model in comparison with other models. The average performance gain in all tasks between our MedMAE and original MAE is approximately $8\%$. From the conducted experiments, we should that models pre-trained with self-supervised learning generalize better to medical tasks than the models pre-trained with supervised learning. Additionally, if the model is pre-trained using medical images, its performance on other medical tasks is much better than a model pre-trained using natural images.

The future direction of this research is to improve the generalization of MedMAE and allow the model to perform multiple medical imaging tasks without having a separate model for each task. This can be done using an efficient continual learning technique.


\bibliographystyle{IEEEbib}
\bibliography{IEEEabrv,Template}

\end{document}